\documentclass[apj,graphicx]{emulateapj}
\usepackage{apjfonts}

\usepackage{graphicx, natbib}

\begin{document}

\shortauthors{Cackett \& Miller}
\shorttitle{Winds and neutron star iron lines}

\title{Broad iron lines in neutrons stars: dynamical broadening or wind scattering?}

\author{Edward~M.~Cackett \altaffilmark{1}}
\author{Jon M. Miller\altaffilmark{2}}

\email{ecackett@wayne.edu}

\affil{\altaffilmark{1} Wayne State University, Department of Physics \& Astronomy, 666 W. Hancock St., Detroit, MI 48201, USA}
\affil{\altaffilmark{2} University of Michigan, Department of Astronomy, 500 Church Street, Ann Arbor, MI 48105, USA}

\begin{abstract}

Broad iron emission lines are observed in many accreting systems from black holes in AGN and X-ray binaries to neutron star low-mass X-ray binaries.  The origin of the line broadening is often interpreted as due to dynamical broadening and relativistic effects.  However, alternative interpretations have been proposed, included broadening due to Compton scattering in a wind or accretion disk atmosphere.  Here we explore the observational signatures expected from broadening in a wind, in particular that the iron line width should increase with an increase in the column density of the absorber (due to an increase in the number of scatterings).  We study the data from three neutron star low-mass X-ray binaries where both a broad iron emission line and absorption lines are seen simultaneously, and show that there is no significant correlation between line width and column density.  This favors an inner disk origin for the line broadening rather than scattering in a wind.

\end{abstract}


\section{Introduction}

X-ray emission from compact objects can often be characterized as a
mixture of two direct emission components: thermal emission from an
accretion disk (and stellar surface, in the case of neutron stars),
and non-thermal emission that often takes a power-law form.  The
non-thermal or ``hard" emission is attributed to a ``corona".  A short
and incomplete list of viable models for the corona includes: thermal
Comptonization in the innermost accretion flow \citep[e.g.][]{sunyaevtitarchuk80},
magnetic flares above the inner disk \citep[e.g.][]{beloborodov99}, 
and a mixture of processes in the base of a relativistic jet \citep[e.g.][]{markoff01}.

Regardless of the specific form of the corona, however, hard X-ray
irradiation of the inner accretion disk is inevitable, and the
reaction or ``reflection" spectrum has been calculated in detail
\citep[e.g.][]{georgefabian91}.  The most prominent feature of
reflection is a broad iron emission line.  Theoretical treatments of
disk reflection have advanced continually, largely driven by
observational results.  Spectroscopic features consistent with the
X-ray irradiation of the inner accretion disk have now been observed
in stellar-mass black holes, massive black holes in active galactic
nuclei, and X-ray binaries harboring neutron stars \citep[for reviews and
summary papers, see, e.g., ][]{miller07, nandra07,cackett10}.

Disk reflection spectra are calculated in the fluid frame of the disk,
and must be convolved with a relativistic blurring function to fit the
spectra that are observed \citep[for recent examples, see][]{brenneman06,dauser10}.  
The full power of disk reflection is then harnessed:
the extent to which the spectra are blurred by Doppler shifts in the
inner disk, and gravitational redshifts near to the compact object,
can be used to measure the inner radius of the disk and black hole
spin \citep[e.g.][and references therein]{miller09, brenneman06, walton13}.  In neutron star X-ray binaries, the inner extent of the disk
can be used to set a limit on the stellar radius \citep[e.g.][]{cackett08,cackett10}, 
and magnetic field \citep{cackett09,papitto09,miller11}.

The basic framework of these measurements and models, and related
assumptions about the geometry of the inner accretion flow, have been
subjected to rigorous tests and potential alternative interpretations
\citep[see e.g.,][]{misra99,reynolds00,lmiller09,reynolds09}.  The recent detection of
reverberation lags in time-resolved spectra of Seyferts is a strong
confirmation of disk reflection spectroscopy in massive black holes \citep[e.g.][]{fabian09,emma11,zoghbi12,zoghbi13, demarco13, cackett13, kara13b, kara13a}.  Reverberation lags have also been seen in a stellar-mass black hole \citep{uttley11} and there are lags in neutron star low-mass X-ray binaries even with hints of reverberation \citep{barret13}.

The fact that reflection is seen in the hard X-ray states of
stellar-mass black holes, in which winds are absent or unimportant
\citep[e.g.][]{miller08,miller12,ponti12}, again signals that the spectra originate
through reflection, not through an alternative process, such as
scattering in a wind \citep[e.g.][]{laurent07,titarchuk09, sim08,sim10a,sim10b}.

The case of neutron stars may be more complicated.  Whereas the
absence of a strong corona in black holes prevents strong disk
reflection, hot blackbody emission from the stellar surface or boundary layer can drive
disk reflection in neutron star low-mass X-ray binaries \citep{cackett10,dai10}.  Thus, 
reflection can occur in much softer states in neutron
stars, and potentially when winds are important.  The question arises,
then, if scattering in a wind or disk atmosphere might shape broad
lines in neutron stars, rather than dynamics in the inner accretion
disk.  Based on recent observations of GX~13+1, for instance,
\citet{diaztrigo12} suggest a wind origin for the breadth of
iron emission lines in neutron stars as a class.  In this work, we critically
examine the properties of disk winds and atmospheres, and broad iron
emission lines, in three neutron star low-mass X-ray binaries.

\subsection{Observational signatures of broadening of lines through scattering in a wind}

To assess whether disk winds are the source of line broadening in neutron star low-mass X-ray binaries we need to consider the observational signatures of such a scenario.  As a photon passes through an accretion disk wind or corona, it will be Compton scattered, leading to a shift in energy of the photon, and hence a broadening of any emission line feature.  The total number of scatterings determines the overall shift in energy for a given photon.  Hence, an increase in the average number of scatterings per photon leads to an increase in the breadth of an emission line.  The number of scatterings is set by both the electron temperature and the optical depth of the wind/corona.  Therefore, for a given electron temperature, an increase in the optical depth will correspond to an increase in the width of an emission line.  Observationally, then, we could expect to see an increase in the measured emission line width ($\sigma$) with an increase in the measured warm absorption column density, when both emission and absorption lines are observed simultaneously.  Such a correlation is claimed in GX 13+1 by \citet{diaztrigo12}, and we study this further, below.

\citet{diaztrigo12} also claim that an increase in the emission line equivalent width with a corresponding increase in the equivalent width of the absorption line (or similarly the warm absorption column density) is a signature that the broad line width is determined by scattering.  However, if the broad emission line flux is nearly constant and absorption increases, this reduces the continuum flux and hence leads to the equivalent width of the emission and absorption lines correlating.  Hence, such a correlation is not a clear observational signature of scattering.  We discuss the strength of this effect in more detail below.

Here, we use values of the emission line width and absorption column density from the literature for 3 objects where both an Fe K emission line and Fe absorption lines are observed simultaneously.  

\begin{deluxetable}{cccc}
\tablecolumns{4}
\tablewidth{0pc}
\tablecaption{Fe emission line width and equivalent hydrogen column density. }
\tablehead{Object & Emission line & $N_{\rm H}$ & Reference \\
                 & width, $\sigma$ (keV) & ($10^{22}$ cm$^{-2}$) &  }
\startdata
 GX~13+1 &    $0.88\pm0.10$ & $17.2\pm8.5$ & \citet{diaztrigo12} \\
 	          &    $0.77\pm0.19$ & $6.6\pm3.0$ & \\
	          &    $0.72\pm0.10$ & $6.6\pm1.5$ & \\
	          &    $0.78\pm0.13$ & $6.0\pm4.3$ & \\
                    &    $0.77\pm0.07$ & $7.1\pm1.4$ & \\
4U~1323$-$62 & 	$0.85\pm0.21$ & $3.6\pm1.0$ & \citet{boirin05} \\
 		& $0.55\pm0.32$ & $6.0\pm3.5$ &  \\
		&  $0.47\pm0.32$ & $14.0\pm8.5$ & \\
4U~1624$-$490 & 	$0.50\pm0.23$	 & $42.3\pm38.2$ & \citet{xiang09} \\
                & $0.18\pm0.06$  & $26.7\pm15.7$ & 
\enddata
\label{tab:sigma_nh}
\end{deluxetable}

\section{Comparison of line width and absorption column}

We searched the literature for neutron star low-mass X-ray binaries that display both a broad iron emission line and iron absorption lines seen simultaneously.  We also require multiple measurements of the line width and absorption column density in order to look for correlations between the two parameters.  This resulted in data from 3 objects: GX~13+1, 4U~1323$-$62 and 4U~1624$-$490.  GX~13+1 is one of only two neutron star low-mass X-ray binaries where significantly blue-shifted absorption lines have been detected \citep{ueda04,diaztrigo12}, while 4U~1323$-$62 and 4U~1624$-$490 are both well-known dipping sources \citep[e.g.][and references therein]{boirin05,xiang09} where their near edge-on inclination leads to strong dips in the lightcurve due to absorbing material close to the accretion disk \citep[e.g.,][]{frank87}.  Note that GX~13+1 is also likely a dipping source \citep{diaztrigo10} and so these two sources are a good comparison.  Sources need to be close to edge-on for both emission and absorption lines to be observed.

\begin{figure}
\centering
\includegraphics[width=8.5cm]{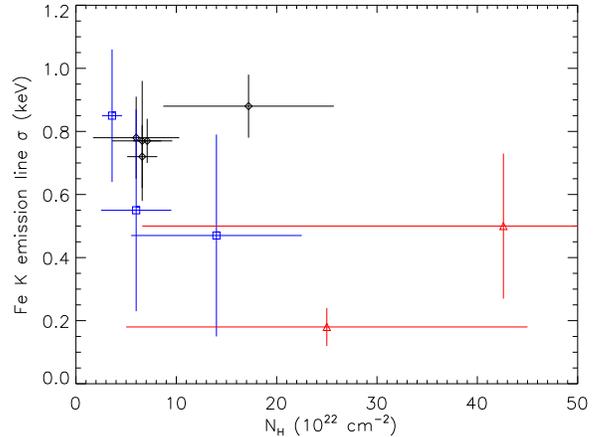}
\caption{Equivalent hydrogen column density for the warm absorber ($N_{\rm H}$) versus iron emission line width ($\sigma$) for GX 13+1 (black, diamonds), 4U~1323$-$62 (blue, squares) and 4U~1624$-$490 (red, triangles).}
\label{fig:sigvscol}
\end{figure}

\begin{figure}
\centering
\includegraphics[width=8.5cm]{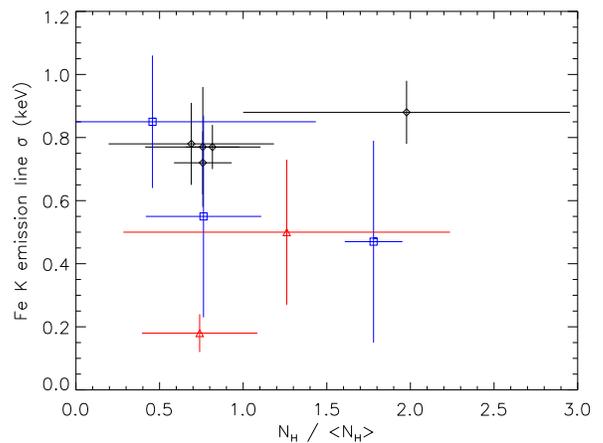}
\caption{Normalized equivalent hydrogen column density for the warm absorber, $N_{\rm H} / \langle N_{\rm H}\rangle$, versus iron emission line width ($\sigma$) for GX 13+1 (black, diamonds), 4U~1323$-$62 (blue, squares) and 4U~1624$-$490 (red, triangles).}
\label{fig:sigvsmeancol}
\end{figure}

We summarize the data we use in Table~\ref{tab:sigma_nh}, here we take the data directly from the papers, yet make the error bars symmetrical by taking the average (note that this does not affect the correlation tests which do not use the errors bars).  For 4U~1624$-$490, \citet{xiang09} fit two absorbers, one warm and one hot.  Here, we sum the two column densities to determine the total absorbing column, taking the mid-point of the uncertainty range.  In Figure~\ref{fig:sigvscol} we plot the absorbing column for the warm absorber versus the iron emission line width for these three sources.  To be able to more directly compare the three objects, in Figure~\ref{fig:sigvsmeancol} we normalize the column densities by the mean column density for each object.  As can be seen from these figures, there is no overall trend of increasing emission line width with increasing column density.  We test this with a Spearman's Rank correlation.  Comparing all the column densities and line widths gives $r=-0.49$ (a weak anti-correlation), which corresponds to a 15\% probability that the anti-correlation is not significant.  If we instead compare the normalized column densities versus line width, we get  $r= -0.14$, which corresponds to a 71\% probability that the correlation is not significant.  Hence, there is no significant correlation when comparing all three sources.

The only source where there are enough data points to test for a correlation individually is GX~13+1.  The Spearman's Rank correlation for GX~13+1 is $r=0.29$ which corresponds to a 64\% probability of no significant correlation.  We further look to see if there is a significant correlation in GX~13+1 by fitting a straight line to the data, using errors in both parameters (utilizing the fitexy routine in IDL).  This leads to a best-fitting line with a slope of $0.012\pm0.016$, consistent with constant emission line width at the 1$\sigma$ level.

We therefore conclude that there is no significant correlation between emission line width and absorbing column density either globally for the three sources or for GX~13+1 alone.

\section{On correlations between emission line equivalent width and absorption column density}

Any increase in the absorption column density will lead to stronger absorption lines but also change the continuum flux.  If the absorption is strong enough it will effect the continuum level through the Fe K band and hence change the measured equivalent width of the Fe K emission line even if the emission line flux remains constant.  Thus, comparing absorption column density (or absorption line equivalent width) with the emission line equivalent width could lead to false correlations.

In order to quantify this effect, we use a model for warm absorption and change the column density to see how it affects the measured equivalent width of an Fe K emission line with constant line flux.  We used an XSTAR grid of models calculated with $n=10^{13}$ cm$^{-3}$ and a turbulent velocity of 300 km/s.  We  also assumed the average luminosity for the source with an irradiating spectrum comprised of a blackbody with $kT = 2$ keV.  We use XSTAR directly rather than using the warmabs model within xspec so that we can calculate the re-emission spectrum also (see next section).  

\begin{figure}
\centering
\includegraphics[width=8.5cm]{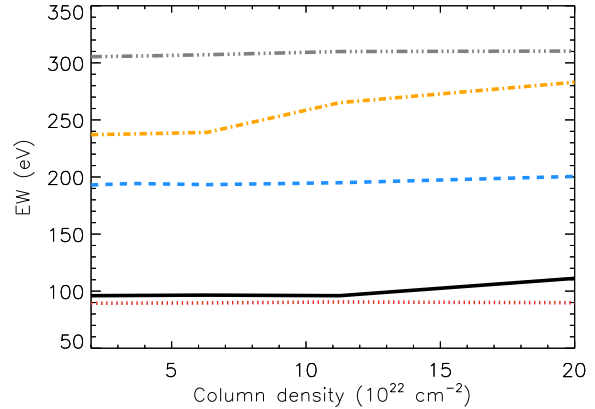}
\caption{Emission line equivalent width vs warm absorber column density for the continuum and line parameters found in GX~13+1 by \citet{diaztrigo12}.  Each line represents the continuum/line parameters for different observations:  observations 4 (gray, dash-triple dotted line), 6 (black, solid line), 7 (red, dotted), 8 (blue, dashed) and 9 (orange, dash-dotted).  A clear increase in emission line EW with increasing column density is seen for 2 of the 5 continuum/line models.}
\label{fig:warmabs}
\end{figure}

We adopt the continuum and emission line parameters found for GX~13+1 by \citet{diaztrigo12} (their table 4) and use our XSTAR model for the warm absorber component.  We change only the column density of the warm absorber and measure the equivalent width of the emission line at each value of $N_{\rm H}$.  In Figure~\ref{fig:warmabs} we show how the emission line equivalent width changes with $N_{\rm H}$ for the continuum and emission line parameters used in each of the 5 observations of GX~13+1 considered \citet{diaztrigo12}.

\begin{figure}
\centering
\includegraphics[angle=270, width=8.5cm]{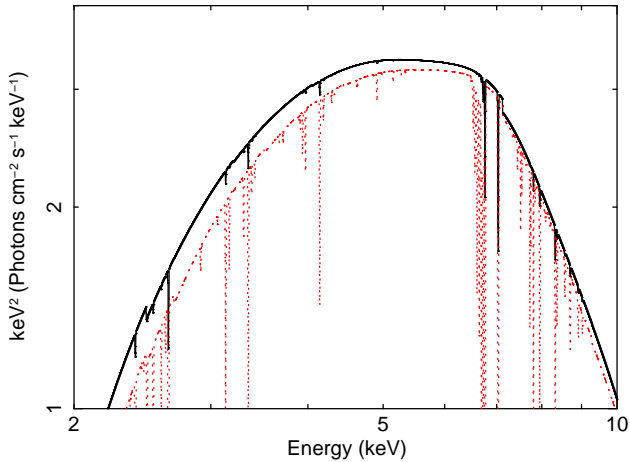}
\caption{The model for observation 9 from \citet{diaztrigo12} with the column density of the warm absorber set to $2\times10^{22}$ cm$^{-2}$ (solid, black line) and $2\times10^{23}$ cm$^{-2}$ (dotted, red line).  A clear change in continuum strength through the Fe K band is seen.}
\label{fig:warmabs_mo}
\end{figure}

The results show that a change in the measured emission line equivalent width with increasing $N_{\rm H}$ depends on the continuum parameters.  However, two of the five models show a clear overall increase in EW when $N_{\rm H}$ increases from $2\times10^{22}$ cm$^{-2}$ to $2\times10^{23}$ cm$^{-2}$.  While several change by just a few eV, the size of the increase in the two clearest is 46 eV and 15 eV.  This can explain some but not all of the change in EW observed in GX 13+1 and intrinsic changes must also be involved.  However, it highlights that a correlation between absorption column density or absorption line equivalent width and emission line equivalent width does not imply that the line is broadened by scattering and can arise simply due to the change in continuum caused by a change in absorption.  We show the change in the model for observation 9 of \citet{diaztrigo12} if we assume a warm absorber with $N_{\rm H} = 2 \times 10^{22}$ cm$^{-2}$ compared to $N_{\rm H} = 2\times 10^{23}$ cm$^{-2}$ in Figure~\ref{fig:warmabs_mo} where a clear change in the continuum strength can be seen.

Furthermore, we can test the statistical significance of the correlation between absorption line EW and emission line EW or between column density and emission line EW in GX 13+1. The strongest absorption line, and the only absorption line to be detected in all 5 observations considered in \citet{diaztrigo12} is Fe XXVI K$\alpha$.  Comparing the Fe XXVI K$\alpha$ absorption line EW with the Fe emission line EW (their table 3) we  get $r = 0.60$ and a 28\% probability that the correlation is false.  If we instead compare the warm absorber column density and Fe emission line EW (their table 4) we get $r =0.66$ (22\% probability).  In other words, there is no statistically significant correlation.

\section{On the emission spectrum from a wind}

\begin{figure}
\centering
\includegraphics[angle=270, width=8.5cm]{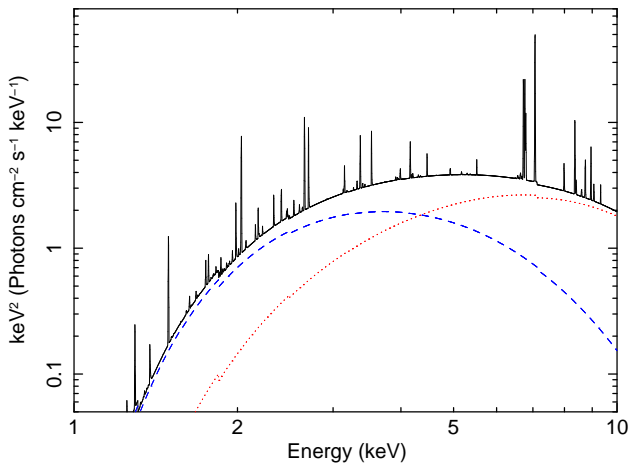}
\caption{Spectrum re-emitted by an accretion disk wind/atmosphere.  This is based on assuming n = $10^{13}$ cm$^{-3}$ and using the warm absorber parameters for observation 6 in \citet{diaztrigo12}.  We also assume the continuum model for observation 6 (disk blackbody is shown by a blue dashed line while the blackbody is shown by a red dotted line) .  The relative normalization of the continuum and emission line spectrum depends on the specific geometry.  Here we have set it so that the 6.7 keV Fe line has the same line flux as quoted in Table 4 of \citet{diaztrigo12}.}
\label{fig:windemis}
\end{figure}

X-ray flux absorbed in the accretion disk wind/corona will be re-emitted, leading the emission lines.  We show an example of such an emission spectrum in Fig.~\ref{fig:windemis}, where we use the same XSTAR models calculated above to determine the emission spectrum.   We used the warm absorber parameters for observation 6 in \citet{diaztrigo12}.  The relative normalization of the continuum and emission line spectrum depends on the specific geometry.  Here we have set it so that the 6.7 keV Fe line has the same line flux as quoted in table 4 of \citet{diaztrigo12}.  Intrinsically narrow emission lines are produced across the spectrum.

Although the emission lines are intrinsically narrow, they will be broadened by Comptonization as well as due to Doppler broadening at the velocities in the outer disk.  Here, we consider whether Compton broadening in a disk atmosphere is sufficient to give the broad lines observed in GX~13+1.  We want to consider what are the realistic conditions in the disk atmosphere/corona and what broadening that will lead to.  The change in energy per Compton scattering goes like $dE/E = kT_e / m_e c^2$, thus it is important to determine a realistic electron temperature.   One of the most detailed studies of an extended accretion disk corona is that of \citet{jimenezgarate05} who apply detailed photoionization models to a high quality {\it Chandra} HETGS spectrum of the neutron star LMXB Her~X-1.  They find a temperature of $kT = 7$ eV ($8\times10^4$ K) based on the Ne IX line and narrow lines $\Delta v \leq 260$ km s$^{-1}$.  A much more extreme example is the black hole X-ray binary GRO~J1655$-$40 which showed a powerful outflowing wind.  The electron temperature there is determined to be $(0.2 - 1.0) \times 10^6$ K \citep{miller06b}.  If we conservatively assume that $kT$ is factor of 14 higher than Her~X-1 and comparable to the extreme case of GRO~J1655$-$40, say $0.1$~keV, then $dE/E = 2\times10^{-4}$, which for a 6.7 keV emission line is $1.3\times10^{-3}$ keV per scatter.     We can calculate the optical depth, $\tau$,  of the warm absorber from the column density via $\tau = N \sigma_T$, where $\sigma_T$ is the Thomson cross-section $= 6.65\times10^{-25}$ cm$^2$.  The largest warm absorber column density for GX~13+1 measured in \citet{diaztrigo12} is $1.7\times10^{23}$ cm$^{-2}$, thus, we get $\tau = 0.11$.  The optical depth is therefore not high enough to give enough scatterings to lead to a line as broad as seen in GX~13+1.

\section{Discussion}

Recent observations of the neutron star low-mass X-ray binary GX~13+1 have been suggested as evidence that the origin of broad emission lines in neutron star low-mass X-ray binaries is due scattering in disk winds \citep{diaztrigo12}.  Here, we have looked at 3 neutron star low-mass X-ray binaries where both a broad iron emission line and absorption lines have been observed simultaneously.  If the breadth of the iron emission line is set by scattering in the disk wind or corona, then one should expect a correlation between the width of the emission line and the absorbing column -- a larger absorbing column leads to more scattering and hence a broader line.  By comparing these two quantities in three neutron stars we find no statistically significant correlation either globally, or on an individual basis for GX 13+1.  We therefore conclude that the observational evidence does not indicate broadening due to scattering in a disk wind.  

One complication that we are not able to test is that a change in electron temperature in the disk wind/corona will change the emission line width -- the electron temperature sets the energy shift per scatter.  While we cannot account for this in our analysis, we can point to the fact that none of the objects show a significant change in line width between different spectra.

\citet{diaztrigo12} claim that in GX 13+1 an increase in emission line equivalent width is correlated with the absorption line equivalent width and that this implies the broad emission line is due scattering.  Firstly, there is no statistically significant correlation between the equivalent widths, or the emission line equivalent width and column density.  Furthermore, a correlation between equivalent widths does not necessarily imply line broadening as a change in continuum strength due to absorption can lead to a change in equivalent width, though this depends on the continuum shape and warm absorber parameters.  

Another insight into the origin of broad iron lines in neutron star
low-mass X-ray binaries can be gleaned by considering magnetic field
constraints that have recently been achieved in accreting X-ray
pulsars.  Broad, likely relativistic iron lines in SAX J1808.6$-$3658
\citep{cackett09,papitto09} and IGR J17480$-$2446 \citep[in Terzan 5;][]{miller11}
 have placed limits on the inner radial extent of the
accretion disk.  In both cases, assuming that the disk is truncated by
the stellar magnetic dipole gives a field strength equivalent to
estimates based on X-ray timing traces of the inner disk extent.
These examples of correspondence clearly signal that the iron emission
line width is driven by dynamics in the inner disk, not by scattering
in a distant wind.

Further support of reflection in neutron star low-mass X-ray binaries is the correspondence between blackbody flux and Fe K emission line flux in 4U 1705$-$44 \citep{lin10} during the soft state. This demonstrates that during soft states the blackbody component (likely associated with the neutron star boundary layer) can be irradiating the accretion disk leading to reflection.  Softer states in neutron star low-mass X-ray binaries do not necessarily have weaker reflection than harder states, and the source of the X-ray irradiation in soft and hard states is likely different \citep{cackett10,dai10}.  

Another point of comparison is the so-called ``accretion
disk corona'' sources, wherein the line of sight is thought to more
persistently intercept the disk atmosphere.  A good example may be
2S~0921$-$63 as observed with the {\it Chandra}/HETGS.  The spectrum
is extremely hard ($\Gamma \simeq 1.1$), and thus likely the result of
scattering rather than direct emission.  A number of emission lines
are observed, including Fe K, with a measured width of $\sigma \leq
0.005$~keV \citep{kallman03}.  Of course, this line is far
narrower than any that would be plausibly associated with the inner
disk, but it is consistent with theoretical treatments of disk
atmospheres \citep[e.g.,][see below]{jimenezgarate01}.

Her X-1 may be another relevant example to consider in this context.
The disk corona in this system has been studied extensively using
gratings X-ray spectroscopy, and detailed analysis has succeeded in
even constraining the electron number density of the gas
\citep{jimenezgarate05}.  In the ``low" state observed using the
{\it Chandra}/HETGS, for instance, neutral, He-like, and H-like lines
are detected.  Each line has a width less than $\sigma \leq 680~ {\rm
km}~ {\rm s}^{-1}$ \citep[or $\sigma \leq 0.02$~keV;][]{jimenezgarate05}.
As we show here, even an electron density of more than 10 times that seen in 
Her~X-1 is not high enough to give sufficient Compton broadening to
match the Fe K emission line width in GX 13+1 given the observed warm
absorber column densities.

Last, if a dense wind is the fundamental requirement for the
production of broad iron lines, then such lines should also be
detected in the spectra of high-mass X-ray binaries.  \citet{furst11}
present a detailed study of the neutron star HMXB GX 301$-$2.
The absorber is found to approach $N_{H} \simeq 10^{24}~ {\rm
cm}^{-2}$, but the width of the iron emission line is just $\sigma$ =
0.03--0.04~keV.  Similarly, the Fe K$\alpha$ lines in HMXBs 
studied by \citet{torrejon10} are also very narrow (unresolved by
{\it Chandra}).  Thus, both theoretical treatments of disk atmospheres
and observations of accretion disk corona sources show that narrow,
not broad, lines originate there.

In conclusion, we do not find any significant correlation between
broad iron line width and the column density of warm absorbers in neutron
star low-mass X-ray binaries.  Thus, we do not currently find any observational evidence
pointing toward line broadening due to a disk wind.  Further observations and detailed 
photoionization modeling of neutron star LMXBs displaying absorption is needed to further
test the disk wind scenario for iron line broadening.

\acknowledgements
We thank Andy Fabian, Chris Reynolds and Rubens Reis for helpful discussions and comments.


\end{document}